\documentclass[sn-mathphys-num]{sn-jnl}

\usepackage{graphicx}%
\usepackage{multirow}%
\usepackage{amsmath,amssymb,amsfonts}%
\usepackage{amsthm}%
\usepackage{mathrsfs}%
\usepackage[title]{appendix}%
\usepackage{xcolor}%
\usepackage{textcomp}%
\usepackage{manyfoot}%
\usepackage{booktabs}%
\usepackage{algorithm}%
\usepackage{algorithmicx}%
\usepackage{algpseudocode}%
\usepackage{listings}%

\unnumbered

\newcommand*{\captionbf}[1]{{\fontfamily{ptm}\selectfont \textbf{#1}}}
\usepackage{bm}
\usepackage{booktabs}

\newcommand{\LossDA}{L_{\text{DA}}}
\newcommand{\LossAdversary}{L_{\text{Adv}}}
\newcommand{\LossTcell}{L_{\text{TCR}}}

\newcommand{\MLPAdversary}{\textsc{MLP}_{\text{Adv}}}
\newcommand{\MLPTcell}{\textsc{MLP}_{\text{TCR}}}
\newcommand{\AUC}{AUC}
\newcommand{\BaselineT}{\text{BASE-T}}
\newcommand{\AdvT}{\text{ADA-T}}
\newcommand{\PreT}{\text{FINE-T}}
\usepackage{soul}

\begin{document}

\title[Article Title]{Transfer Learning for T-Cell Response Prediction}


\author*[1,3]{\fnm{Josua} \sur{Stadelmaier}}\email{josua.stadelmaier@uni-tuebingen.de}

\author[2]{\fnm{Brandon} \sur{Malone}}

\author[1,4]{\fnm{Ralf} \sur{Eggeling}}

\affil[1]{\orgdiv{Department of Computer Science}, \orgname{University of T\"{u}bingen}, \orgaddress{\street{Sand 14}, \city{T\"{u}bingen}, \postcode{72076}, \country{Germany}}}

\affil[2]{\orgname{NEC OncoImmunity}, \orgaddress{\city{Oslo},  \country{Norway}}}

\affil[3]{\orgdiv{Current affiliation: Quantitative Biology Center (QBiC)}, \orgname{University of T\"{u}bingen}, \orgaddress{\postcode{72076 T\"{u}bingen}, \country{Germany}}}

\affil[4]{\orgdiv{Institute for Bioinformatics and Medical Informatics (IBMI)}, \orgname{University of T\"{u}bingen}, \orgaddress{\postcode{T\"{u}bingen}, \country{Germany}}}


\abstract{We study the prediction of T\nobreakdash-cell response for specific given peptides, which could, among other applications, be a crucial step towards the development of personalized cancer vaccines. It is a challenging task due to limited, heterogeneous training data featuring a multi-domain structure; such data entail the danger of shortcut learning, where models learn general characteristics of peptide sources, such as the source organism, rather than specific peptide characteristics associated with T\nobreakdash-cell response. Using a transformer model for T\nobreakdash-cell response prediction, we show that the danger of inflated predictive performance is not merely theoretical but occurs in practice. Consequently, we propose a domain-aware evaluation scheme. We then study different transfer learning techniques to deal with the multi-domain structure and shortcut learning. We demonstrate a per-source fine tuning approach to be effective across a wide range of peptide sources and further show that our final model is competitive with existing state-of-the-art approaches for predicting T\nobreakdash-cell responses for human peptides.}




\maketitle

\section{Introduction}
The human immune system consists of humoral and cell-mediated mechanisms, with T~cells playing a significant role in the latter. T~cells recognize and eliminate cancerous or infected cells by detecting peptides presented on the cell surface by major histocompatibility complex (MHC) molecules, which can be grouped into two classes, MHC I and MHC II \citep{Murphy2018}.

Peptide-based vaccines are a promising approach for the personalized treatment of cancer \citep{sahin2018personalized} and might allow more precise vaccines against infectious diseases \citep{heitmann2022covid}.
Such vaccines can be developed by selecting peptides that are specific to a pathogen or tumor. Since only a limited number of peptides can be included in the vaccine, their selection should be based on their probability of inducing a T\nobreakdash-cell response. Predicting this probability is an important task in the development of peptide vaccines \citep{sahin2018personalized}.

From a computational perspective, this task can be separated into two sub-tasks: first predicting whether a peptide is presented on the cell surface by an MHC molecule and afterwards predicting whether a presented peptide induces a T\nobreakdash-cell response. For the first task, a large amount of available experimental data has led to accurate machine learning-based predictions \citep{netmhc,mhcflurry,bertmhc}. 
The second task can, despite some attempts~\citep{prime2_0, HLA_CD4_Immunogenicity}, still be considered as an open problem, in part due to fewer experimental data from T\nobreakdash-cell assays~\citep{iedb}. 

Aside from small sample sizes, another challenge arises from peptides in T\nobreakdash-cell response data originating from different sources, such as viruses, bacteria, or human proteins. Moreover, peptides feature patterns that are specific to the allele of the MHC molecule they are presented by \citep{rammensee1995mhc}. Both factors lead to data heterogeneity, which is either unaccounted for by existing approaches \citep{HLA_CD4_Immunogenicity,tcr_landscape} or handled by limiting the model to  selected MHC alleles and considering small peptides, such that positions that are affected by MHC allele-specific patterns can be identified \citep{lee2023robust, prime2_0, mhc_i_tcell}. 

To leverage all the available data, we investigate more flexible approaches that do not limit the selection of peptides based on their source, length, and presenting MHC allele.
Based on our data analysis, we show how the heterogeneity of the resulting data set can be viewed as a multi-domain structure, which motivates the application of different transfer learning techniques in combination with the flexible transformer architecture \citep{transformer}. We further investigate how the prediction performance on individual domains is affected by including data from the other domains in the training process, where improved performance is denoted as positive transfer and reduced performance as negative transfer \citep{RosensteinNegativeTransfer}.

For all empirical studies it is critical to keep in mind that having several underlying domains with varying fractions of positives entails the risk of shortcut learning \citep{li2021deepimmuno, brown2023detecting, geirhos2020shortcut}, that is, using domain-specific features to classify peptides based on their domain identity instead of T\nobreakdash-cell response-specific patterns. To obtain shortcut-invariant model performance estimates, we thus propose a domain-aware evaluation scheme. 

Our results indicate that shortcuts based on peptide sources and MHC alleles are indeed learned by the transformer and lead to inflated performance estimates when the domain-aware evaluation is not applied. We further observe that per-source fine-tuning is effective in enabling positive transfer across a wide range of peptide sources and show that our final model is competitive with existing state-of-the-art approaches for predicting T\nobreakdash-cell responses for human peptides.

\section{Materials and Methods}
We first describe the T\nobreakdash-cell response data set construction and analyze the domain structure in the data. Then follows an introduction of the baseline transformer model for T\nobreakdash-cell response prediction. To account for the identified domain structure, we describe two transfer learning model variants and a special evaluation scheme.

\subsection{Construction of a T\nobreakdash-cell response data set}
We use the Immune Epitope Database\footnote{https://www.iedb.org} (IEDB) \citep{iedb} to construct a T\nobreakdash-cell response data set.
A data point consists of the amino acid sequence of the peptide, a binary label representing the T\nobreakdash-cell response in the form of IFN$\gamma$, the allele of the MHC molecule that presented the peptide to a T~cell, the class of the MHC allele (I or II), and the source of the peptide. The source can be an organism or a virus.

For some peptides in the data set, the MHC alleles are not specified in the IEDB. Additionally, the MHC allele information is incomplete since one peptide can often be presented on several MHC alleles. However, not all combinations of peptides and MHC alleles are tested.
Since the existing MHC information in the IEDB is mostly based on predictions, we use existing models to obtain a list of MHC alleles for each peptide in a consistent way.
NetMHCpan~4.1 \citep{netmhc} serves as a predictor for peptides with MHC class I alleles and NetMHCIIpan~4.0 \citep{netmhc} for peptides with MHC class II alleles.
We select the 100 most frequent alleles, which account for 98.7\% of peptide-MHC combinations in the IEDB data set, and use all MHC class-matched combinations of peptides and MHC alleles as input for the two predictors.

We assign an allele to a peptide if it is predicted to be a weak binder. NetMHCpan~4.1 and NetMHCIIpan~4.0 define a peptide as weak binder if its prediction score is within the top 2\% of prediction scores for random natural peptides.
Peptides that are predicted to not bind to any of the alleles get assigned a default allele for the respective MHC class. This applies to 29.7\% of class I peptides and 66.5\% of class II peptides.
Since the MHC alleles are only relevant for us to identify MHC allele-specific patterns in peptides, the relatively large fraction of default alleles is not problematic because peptides with a default allele likely have no strong allele-specific patterns.
Table \ref{tab:data_stats} shows the resulting statistics about the MHC classes, T\nobreakdash-cell response labels, and peptide lengths of the T\nobreakdash-cell response data.

\begin{table}[ht]
  \caption{Statistics of the data set. The columns for positives/negatives refer to the number of peptides with the respective T\nobreakdash-cell response label.}
  \centering
     \begin{tabular}{lrrrr}
       \toprule
              & MHC alleles & Positives & Negatives & Pep. lengths \\
\midrule
MHC I   & 54          & 4,460      & 12,697   &  8-15\\
MHC II  & 46          & 10,247      & 8207    & 9-25\\
Total         & 100         & 14,707      & 20,904 & 8-25\\
\bottomrule
\end{tabular}
\label{tab:data_stats}
\end{table}

\subsection{Domain structure in T\nobreakdash-cell response data}
\label{sec:domain_structure}
\begin{figure*}[ht]
  \centering
  \includegraphics[width=0.80\textwidth]{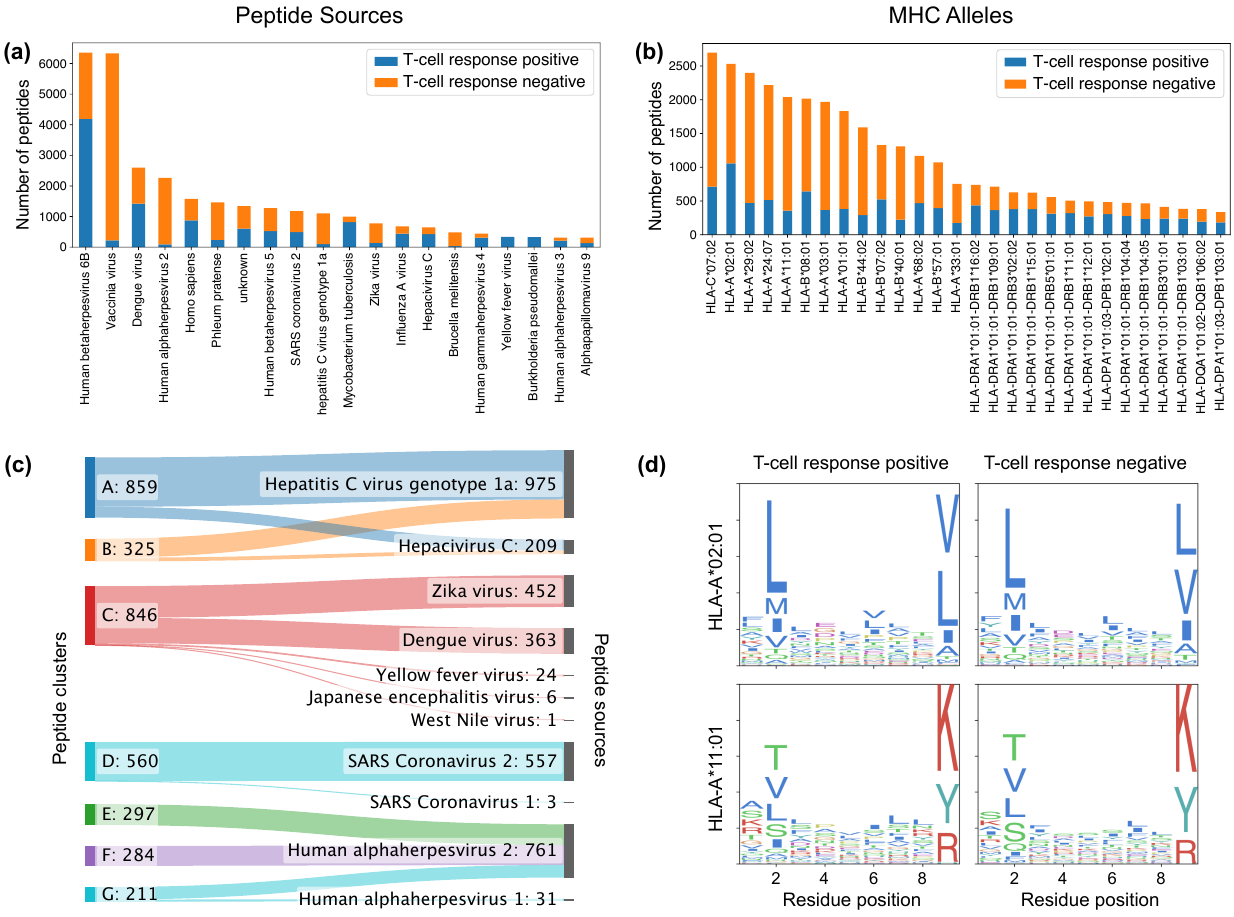}
  \caption{Domain structure of T-cell response data.
  \captionbf{(a)} Distribution of T-cell response positives and negatives per peptide source.
  \captionbf{(b)} Same plot for MHC alleles.
  \captionbf{(c)} Clusters (indexed with A-G) of similar peptides (left) and the sources of peptides the clusters consist of (right). Numbers show peptide counts.
  \captionbf{(d)} Sequence logos for the MHC alleles HLA-A*02:01 (first row) and HLA-A*11:01 (second row). The columns represent T-cell response positives (left) and negatives (right).
  }
  \label{fig:data_domain_structure}
\end{figure*}

Based on the data analysis presented in Figure \ref{fig:data_domain_structure}, we identify two domain structures in the T\nobreakdash-cell response data set, peptide sources (left column) and MHC alleles (right column). The next paragraphs describe these two domain structures in more detail. 

Figure~\ref{fig:data_domain_structure}a shows that the peptides in the T\nobreakdash-cell response data set originate from several sources and that the fraction of T\nobreakdash-cell response positives varies strongly between sources. 
When studying T\nobreakdash-cell responses for one peptide source, a protein of that source is often screened by testing a set of largely overlapping peptides that originate from the protein. 
These overlaps lead to clusters of peptides that make peptides within one source more similar than peptides from a different source.

Motivated by the size of the binding core in peptides \citep{data_split}, we define that peptides correspond to the same cluster if they share a sub-sequence of length nine. The Sankey diagram in Figure~\ref{fig:data_domain_structure}c confirms that the seven largest peptide clusters typically consist of peptides from only a few sources.

In analogy to the distribution of peptide sources in Figure~\ref{fig:data_domain_structure}a,
Figure~\ref{fig:data_domain_structure}b shows the distribution of MHC alleles. As peptides can be associated with several MHC alleles, we show a subset of MHC alleles that together cover a large fraction of peptides. 
Compared to peptide sources, there is a lower variability in the fraction of T\nobreakdash-cell response positives between MHC alleles.

The distinct binding motifs of MHC alleles \citep{rammensee1995mhc} introduce an MHC-based domain structure in T\nobreakdash-cell response data. We visualize the sequence logos for two exemplary MHC alleles and peptides of length nine, separately for T\nobreakdash-cell response positives and negatives, in Figure~\ref{fig:data_domain_structure}d.
At the anchor positions two and nine, there are obvious differences between the MHC alleles in the sequence logos. 
However, comparing sequence logos between T\nobreakdash-cell response positives and negatives does not reveal obvious differences.

Since studies of T\nobreakdash-cell responses often focus on just one peptide source and one MHC class or only consider peptides that are presented on certain MHC alleles, the joint distribution of peptide sources and MHC alleles is not uniform (see Supplement S1). Consequently, peptide sources can also feature the patterns of the MHC alleles with which they co-occur.

\begin{figure}[ht]
  \centering
  \includegraphics[width=.40\textwidth]{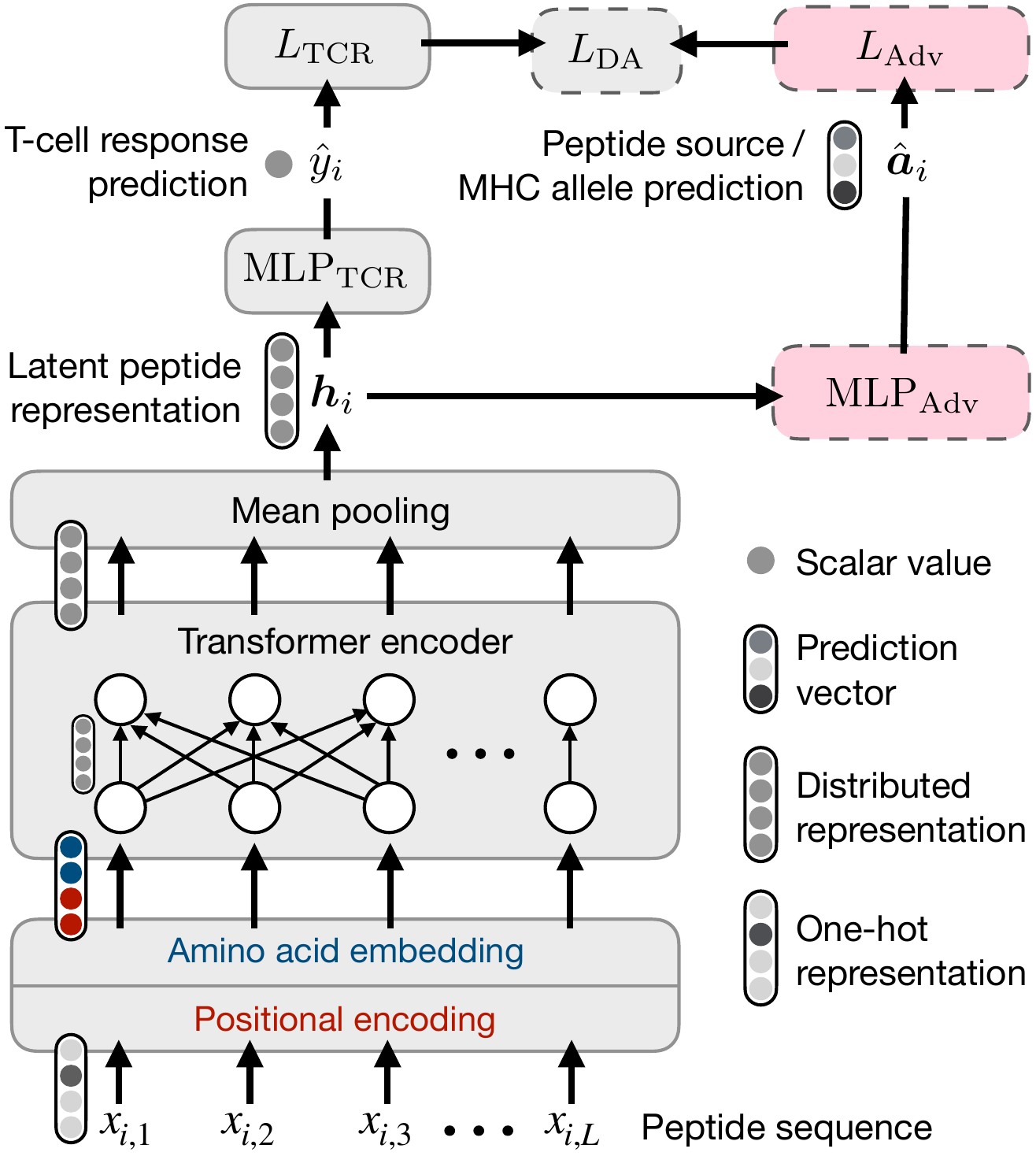}
  \caption{Model architecture for T-cell response prediction. Shading of boxes indicates with which objectives the components are trained. Boxes with dashed borders are only used in the adversarial domain adaptation setting.}
  \label{fig:architecture}
\end{figure}

\subsection{Transformer model}
\label{sec:transformer_model}
We use the transformer architecture \citep{transformer}, as shown in Figure~\ref{fig:architecture}, to capture T\nobreakdash-cell response-specific patterns in peptide sequences.
To apply the transformer, we formalize the data set to consist of data points $\{(\bm{x}_i, y_i)\}_{i=1}^{N_{\text{data}}}$, where $\bm{x}_i = (x_{i,1}, \dots, x_{i,L})$ is the $i$-th peptide sequence with each of the $x_{i,l}$ representing one of 20 amino acids or a padding token. The padding token allows the sequences to all have the same length $L$, which corresponds to the maximal peptide length.
We denote the binary label indicating whether the peptide $\bm{x}_i$ leads to a T\nobreakdash-cell response with $y_i$.

Our model gets a one-hot representation of the peptide sequence $\bm{x}_i$ as input. Then the model maps each amino acid or padding token $x_{i,j}$ of the peptide sequence to a learned embedding $\bm{e}_{i,l} \in \mathbb{R}^{d/2}$.
To tell the transformer the positions at which the amino acids occur in the peptide, we define a learned embedding $\bm{p}_l \in \mathbb{R}^{d/2}$ for each position $l \in \{1, \dots, L\}$. Additionally, we use positional encodings $\bm{s}_l \in \mathbb{R}^{d/2}$ based on sine functions with varying frequencies (see \cite{transformer} for more details).
We then add the two variants of positional vectors and concatenate them with the amino acid embeddings to obtain a vector representation
$\left(\begin{smallmatrix} \bm{e}_{i,l} \\ \bm{p}_l + \bm{s}_l \end{smallmatrix}\right) \in \mathbb{R}^{d}$ for each element $x_{i,l}$ of the peptide sequence $\bm{x}_i$.

The resulting sequence of vector representations is the input of a transformer encoder layer, which updates the vector representations based on interactions between amino acids using a mechanism called multi-head self-attention (see \cite{transformer} for more details).

As the output of the transformer encoder is again a sequence of vector representations, we apply mean pooling to obtain a single vector $\bm{h}_i \in \mathbb{R}^{d}$, which is a latent representation of the $i$-th peptide.
In a similar way as described here, the transformer architecture has been used in earlier work to learn representations of peptides \citep{bertmhc} and proteins \citep{tape}.

Using the latent peptide representation $\bm{h}_i$ as input, a multilayer perceptron (MLP) with one hidden layer of size $d$ computes the T\nobreakdash-cell response prediction $\hat{y}_i = \MLPTcell(\bm{h}_i)$.

All described components are trained to minimize the loss
\begin{equation}
\LossTcell = \frac{1}{N_{\text{data}}} \sum_{i=1}^{N_{\text{data}}} \text{CE}(y_i, \hat{y}_i),
\end{equation}
with the binary cross-entropy between T\nobreakdash-cell response labels and predictions being defined as
\begin{equation}
\text{CE}(y_i, \hat{y}_i) = -(y_i \text{log}(\hat{y}_i) + (1-y_i)\text{log}(1-\hat{y}_i)).
\end{equation}

\subsection{Adversarial domain adaptation}
\label{sec:adv_domain_adaptation}
As described in Section \ref{sec:domain_structure}, the T\nobreakdash-cell response data set has an underlying domain structure of peptide sources and MHC alleles.
This heterogeneity suggests that predictions might be biased by the class imbalance (see Figures \ref{fig:data_domain_structure}a \ref{fig:data_domain_structure}b). Additionally, domain-specific features might be used by a machine learning model instead of features that generalize over MHC alleles and peptide sources.

To address these issues, we apply a transfer learning technique called adversarial domain adaptation \citep{tzeng2017adversarial}, to encourage the model to perform predictions independently of the domain identity of a peptide. Unlike the typical domain adaptation setting with a large amount of training data for a source domain and no or little training data for the target domain \citep{survey_transfer_learning}, in our setting, there are more than two domains and all domains can act as target and source domain at the same time.
We use a variant of adversarial domain adaptation that is adapted to our setting. This variant is very similar to \emph{multi-domain adversarial learning} \citep{multi_domain_adv_learning}.

The model components for implementing adversarial domain adaptation are marked with dashed borders in Figure~\ref{fig:architecture}.
To encourage domain invariance, we define a new training objective:
\begin{equation}
\LossDA = \LossTcell - \lambda \LossAdversary.
\end{equation}
As before, minimizing $\LossTcell$ is necessary for performing the main task, T\nobreakdash-cell response prediction.
The term $-\LossAdversary$ measures how much information about the domain identities is encoded in the latent peptide representations $\bm{h}_i$.
The strength of domain adaptation is determined by the hyperparameter $\lambda$.
In the following, we explain adversarial domain adaptation considering peptide sources as domains.
To capture how much information about peptide sources is encoded in $\bm{h}_i$, we train an adversarial $\MLPAdversary$ with input $\bm{h}_i$ and output $(\hat{a}_{i,1}, \dots, \hat{a}_{i,N_{\text{source}}})^{\intercal} = \hat{\bm{a}}_i \in \mathbb{R}^{N_{\text{source}}}$ to predict how likely peptide $i$ originates from each of the $N_{\text{source}}$ peptide sources.

$\MLPAdversary$ is trained concurrently with the other parts of the T\nobreakdash-cell response model using the following multi-label classification objective $\LossAdversary$, where $a_{i,j}$ are the ground truth peptide source labels:

\begin{equation}
\LossAdversary = \sum_{i=1}^{N_{\text{data}}} \; \sum_{j=1}^{N_{\text{source}}} \text{CE}(a_{i,j}, \hat{a}_{i,j}).
\end{equation}

While the parameters of $\MLPAdversary$ are trained to minimize the peptide source classification error, $\LossAdversary$ is re-used with a negative sign in $\LossDA$. This means that latent peptide representations produced by the transformer are encouraged to confuse $\MLPAdversary$ by making latent representations from different domains more similar, suggesting the name \emph{adversarial} domain adaptation.
The above explanations for peptide sources can be directly generalized to MHC alleles.

\subsection{Per-source fine-tuning}
When one machine learning model is trained on data from several domains, the performance on individual domains can be reduced compared to models that are trained on data from one domain only. Such instances of negative transfer are more likely to occur if domains are too dissimilar \citep{RosensteinNegativeTransfer}.
Negative transfer is also possible when applying adversarial domain adaptation because there is only one model for all domains. 
Additionally, data points from all domains are encouraged to be represented similarly in the model.
To avoid negative transfer, a transfer learning technique is needed that provides more flexibility to adapt to individual domains, while still allowing positive transfer from other domains.

As an alternative to adversarial domain adaptation, we propose to pre-train a T\nobreakdash-cell response model until convergence on all domains and then continue training on only one domain. This second step is also called fine-tuning. We allow all model parameters to be updated during fine-tuning such that the model has the necessary flexibility to correct instances of negative transfer.
This transfer learning technique is commonly applied in combination with the transformer and proteins \citep{tape} or peptides \citep{bertmhc}.

\subsection{Bag of amino acids baseline}
To allow a direct comparison of our proposed transformer-based models to a simpler model that is trained on the same data set, we use a \emph{bag of amino acids} (Bag-Of-AA) baseline, since it can be viewed as a simplified version of PRIME~\citep{prime2_0}.

This baseline model counts how often each of the 20 amino acids occurs in a given peptide and puts the frequencies into a 20-dimensional vector. An MLP with one hidden layer receives these vectors as input and is trained to predict the T\nobreakdash-cell responses. As with the transformer models, we use the binary cross-entropy as training objective.
The Bag-Of-AA model does not consider positional information, that is, it does not see at which positions the amino acids occur in the peptide. The amino acid frequencies allow the model to capture possible preferences for certain amino acids that T~cells might have.

\subsection{Model evaluation}
\label{sec:model_evaluation}
We measure the model performance with the area under the ROC curve (AUC) based on the ground-truth T\nobreakdash-cell response labels $y_i$ and the predictions $\hat{y}_i$. 

We partition the data set $\mathcal{D}$ into five disjoint subsets $\mathcal{D}_{1},\dots,\mathcal{D}_{5}$, and then permute and re-combine them into training, validation, and test sets. This setting ensures that the partitions are large enough to each have a representative distribution of the domains. 

Within a peptide cluster, sequences can be nearly identical (cf. Section \ref{sec:domain_structure}). Splitting the data completely at random could lead to such near identical sequences being distributed among training, validation, and test sets. This can potentially cause inflated generalization estimates.
To prevent this, we assign peptide clusters as a whole (instead of single peptides) to a randomly selected partition \citep{data_split}.

Throughout this manuscript, we carry out all method development and hyperparameter tuning on the union of the first four data subsets. Using $\mathcal{D}_1,\dots,\mathcal{D}_{4}$ as input for a four-fold cross validation, we refer to the resulting mean AUC over the four folds as \emph{validation performance}. This leaves $\mathcal{D}_{5}$ as unseen, independent test data set for evaluating generalization capabilities in the final study, where we refer to the evaluation of the predictions on $\mathcal{D}_{5}$ while training on $\mathcal{D}\setminus \mathcal{D}_{5}$ as \emph{test performance}. 

In order to obtain an uncertainty quantification of the test performance, we additionally carry out a test evaluation for each of the other subsets $D_j$ for $j\in \{1,\dots,4\}$. Training on $\mathcal{D}\setminus \mathcal{D}_{j}$ always involves the re-tuning of numeric hyperparameters via four-fold cross-validation, so the procedure can be regarded as a nested cross validation. We note, however, that more general choices regarding the model architecture and the range of tunable hyperparameters originate from the the primary validation study carried out on $\mathcal{D}\setminus \mathcal{D}_5$.  

\subsection{Detection of shortcut learning}
\label{sec:shortcut_evaluation}
The domain structure with different T\nobreakdash-cell response positive/negative ratios per domain (see Section \ref{sec:domain_structure}) suggests that \textit{shortcuts} can be learned to predict T\nobreakdash-cell responses by first picking up domain-specific features to identify domains and then outputting the majority class within the domain.

In our analysis, we evaluate model performance to account for such shortcuts. In particular, we group the evaluation data into subsets such that each subset contains peptides from only one combination of peptide source and MHC allele. 
We further weight by subset size and peptide:MHC assignments to ensure each peptide contributes equally to the overall performance (see Supplement S2 for details).
Since this evaluation setting ensures that performance estimates are not inflated by the use of shortcuts, we refer to this strategy as \emph{shortcut adjusted evaluation}.

\clearpage

\section{Results}
We first evaluate the practical impact of shortcut learning and study the behavior of the adversarial domain adaptation method. Afterwards, we demonstrate the effectiveness of the per-source fine-tuning. Finally, we evaluate that method on independent test data and compare with existing models from the literature. We provide details on the implementation and the model hyperparameters in the Supplement S3.

 \begin{figure}[ht]
   \centering
   \includegraphics[width=0.90\textwidth]{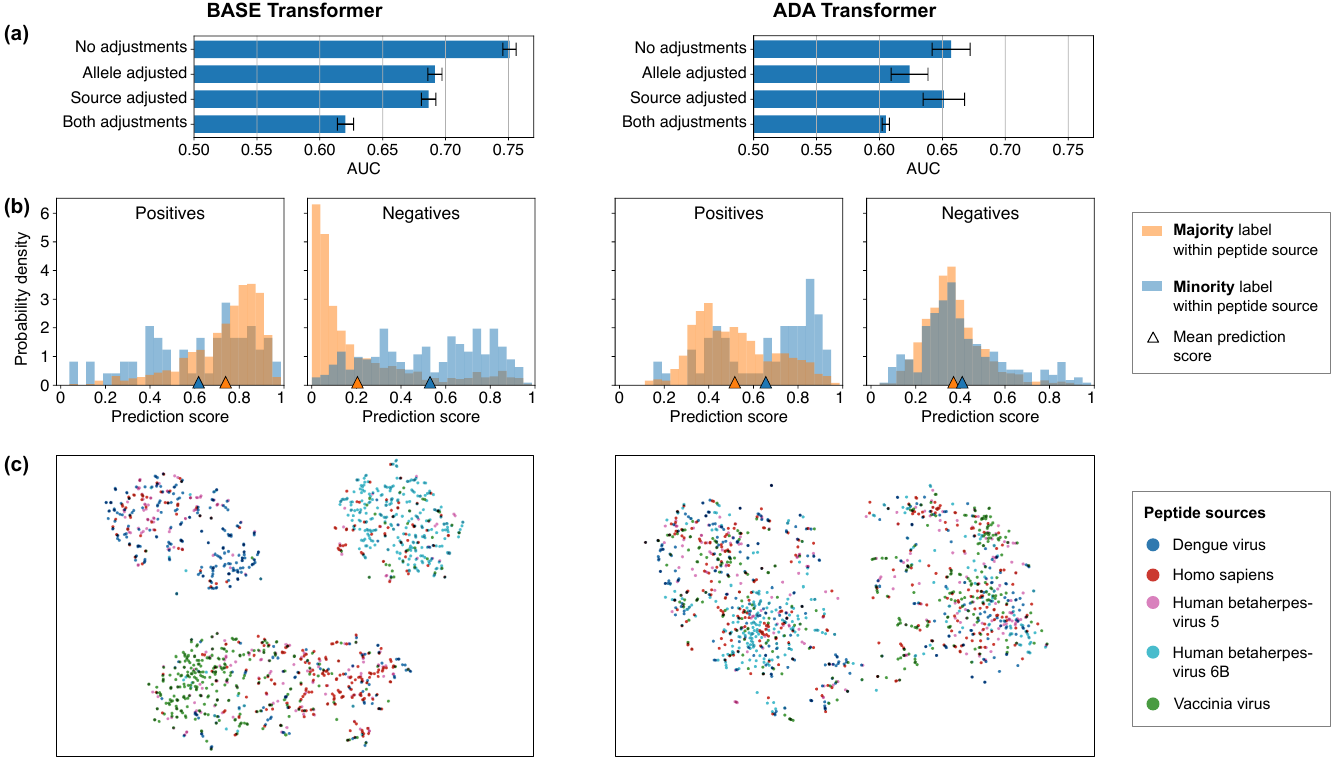}
   \caption{Shortcut learning and the effect of adversarial domain adaptation. The left column shows results for \BaselineT{} and the right column shows corresponding results for \AdvT{} with adversarial domain adaptation being applied on peptide sources.
   \captionbf{(a)} Model performance on validation data with different settings of accounting for shortcuts in the evaluation. For the "allele adjusted" performance, peptides are grouped by MHC alleles in the evaluation to detect shortcuts based on MHC alleles. "Source adjusted" is analogous for grouping peptides by their source.
   \captionbf{(b)} Distribution of prediction scores, separated by the T-cell response labels. For each label, two distributions are shown that correspond to the two most frequent peptide sources, Human betaherpesvirus 6B (majority of labels is positive) and Vaccinia virus (majority of labels is negative). The markers on the x-axis show the mean prediction scores of the two distributions.
   \captionbf{(c)} t-SNE visualization of the latent peptide representations~$\mathbf{h}$.
   } 
   \label{fig:source_domain_adaptation}
 \end{figure}

\subsection{Shortcut learning in practice}
\label{sec:baseline_transformer_results}
We investigate the prevalence of shortcut learning in T\nobreakdash-cell response prediction models trained on the full T\nobreakdash-cell response data set, that is, on all peptide sources and MHC alleles (MHC I+II) combined. In order to study whether shortcut learning can be prevented by encouraging domain invariance in the model architecture and whether this constraint could serve as regularizer to improve the model performance, we compare two different models: the baseline transformer \BaselineT, and \AdvT, the adversarial domain adaptation approach with domain adaptation strength set to $\lambda=10$. The results for both models are summarized side by side in Figure~\ref{fig:source_domain_adaptation}.

The prediction performance in terms of \AUC\, with and without shortcut adjustment in the evaluation is shown in Figure~\ref{fig:source_domain_adaptation}a. 
For \BaselineT\, we observe that the shortcut adjustment w.r.t. alleles and peptide sources leads to an absolute \AUC\, decrease of 0.06 when applied separately, and a further decrease by 0.07 when combined, which indicates that the model learns shortcuts based on both MHC allele-specific and peptide source-specific features.
In contrast, for \AdvT\, the performance without adjustments and with source shortcut adjustment is nearly identical, which demonstrates that adversarial domain adaptation is effective in reducing shortcut learning based on peptide sources. When adjusting for both possible shortcuts in the evaluation, \AdvT{} leads to slightly reduced predictive performance compared to \BaselineT. 

To verify that shortcuts are actually based on the label imbalance within different domains (see Figures \ref{fig:data_domain_structure}a and \ref{fig:data_domain_structure}b), we plot the prediction scores of peptides from the two most frequent peptide sources, Human betaherpesvirus 6B (mostly positives) and Vaccinia virus (mostly negatives) as histograms, separated according to the true response labels in Figure~\ref{fig:source_domain_adaptation}b. 
For \BaselineT, we find that the distribution of prediction scores is always shifted towards the majority class within the peptide source: Human betaherpesvirus 6B peptides with a positive label tend to get higher prediction scores than Vaccinia virus peptides with a positive label. For negatives, the Vaccinia virus peptides get lower prediction scores than Human betaherpesvirus 6B peptides.
\AdvT{} overcorrects the bias towards the majority label within sources for the T\nobreakdash-cell response positives, as the Vaccinia virus peptides (mostly negatives) now get even higher prediction scores than Human betaherpesvirus 6B peptides. For the T\nobreakdash-cell response negatives, the distributions of prediction scores corresponding to the two peptide sources are now well aligned, indicating that no source shortcuts are used. These findings may also explain the slight drop in overall prediction accuracy of \AdvT{} observed in Figure~\ref{fig:source_domain_adaptation}a, as the number of correct predictions in the minority classes are increased at the expense of additional errors in the majority classes.
We also carried out this study for other peptide sources (Supplement S4), where a similar effect is visible for most of them, albeit to a smaller degree, which is likely due to the smaller sample size.

In addition to the analysis of model outputs, we investigate the internal representations of the models regarding indications of shortcut learning and the effect of \AdvT{}. To use shortcuts based on peptide sources, we expect \BaselineT{} to learn source-related features in the latent peptide representations $p$. 

To test whether this is the case, we use t-stochastic neighborhood embedding (t-SNE) by \cite{tsne} to project the high-dimensional latent peptide representations $p$ to two-dimensional points.
We here use the t-SNE implementation of the TensorFlow embedding projector with the default parameters for learning rate and perplexity.

The cluster structure in the t-SNE plot of latent peptide representations in Figure~\ref{fig:source_domain_adaptation}c provides indeed further evidence that features related to peptide sources are learned by \BaselineT{}. For example, the Human betaherpesvirus 6B peptide representations (cyan) concentrate mostly in the top right corner while the Vaccinia virus peptides (green) can be found mostly in the lower left corner. Conversely, the t-SNE plot is less clustered for \AdvT{}, demonstrating that source-related features are removed from the peptide representations.

\subsection{Negative transfer between domains}
Next, we investigate why applying adversarial domain adaptation to peptide sources does not improve performance despite encouraging transfer between the sources. One hypothesis is that peptide sequences from different sources are too dissimilar, which leads to negative transfer.

To test this hypothesis, we compare \BaselineT, being trained all peptide sources and MHC alleles (\textit{multi-domain} setting), with a combination of \BaselineT{} models that eliminate transfer between sources and alleles respectively. All evaluations apply the full shortcut adjustment.

In a \textit{per-source} setting, we train separate \BaselineT{} models for every peptide source, each of which contains peptides from several MHC alleles.
To make the performance of the models in the \textit{per-source} setting directly comparable to the \textit{multi-domain} setting, we first aggregate the predictions for the each source (performed by different models) into one data set again.
Then we apply the evaluation on groups of peptide source/MHC allele combinations exactly as in the \textit{multi-domain} setting.
This evaluation also ensures that only predictions coming from the same per-source model are compared to each other because AUC values are computed per peptide source/MHC allele combination.
Conversely, the \textit{per-allele} setting is a combination of separate \BaselineT{} models for each MHC allele, each of which is trained on data from several peptide sources.

The corresponding validation study (Supplement S5) shows that the \AUC\, of the multi-domain model is reduced by 0.05 compared to the per-source models and by 0.02 compared to the per-allele models, confirming our hypothesis that there is negative transfer between peptide sources and, to a smaller degree, between MHC alleles.

\begin{figure}[ht]
  \centering
  \includegraphics[width=1.\textwidth]{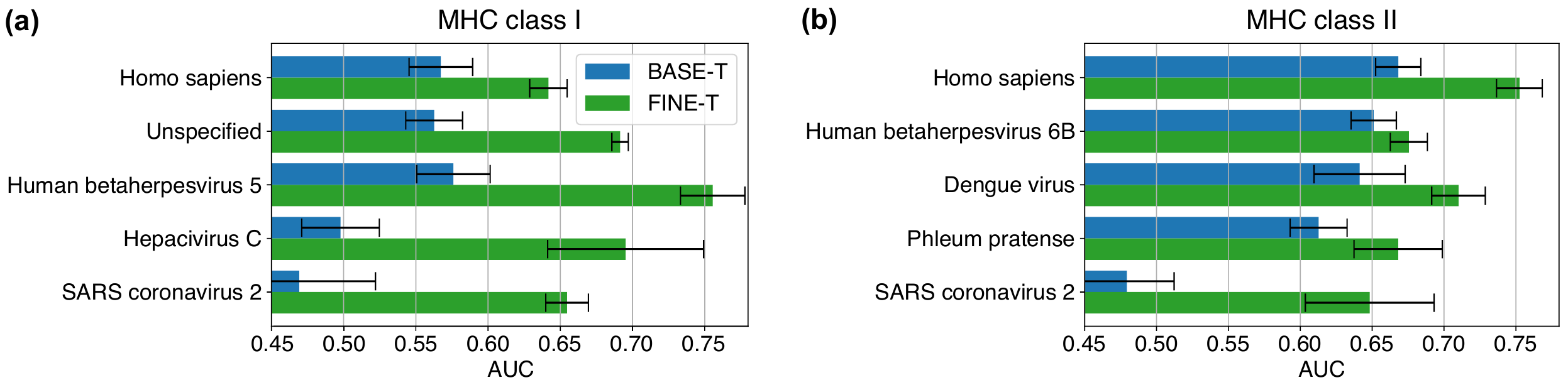}
  \caption{Validation performance of \BaselineT{} and \PreT{} models for several peptide sources. Results for MHC I are shown in \captionbf{(a)} and for MHC II in \captionbf{(b)}. For each MHC class, the five most frequent peptide sources are selected. Peptide sources with only positives or only negatives in one of the test data partitions are excluded.}
  \label{fig:multi_source_valid_base_pre}
\end{figure}
\subsection{Improved predictions by per-source fine-tuning}

Since negative transfer mainly occurs between sources and since per-source models achieve the highest accuracy in predicting T\nobreakdash-cell responses, we explore another transfer learning technique. We initially train \BaselineT{} on all sources and subsequently fine-tune per-source models with model parameters initialized using \BaselineT{}. With this per-source fine-tuning approach, which we refer to as \PreT{}, we aim to combine the advantages of per-source models with the possibility of positive transfer between sources. Since some peptide sources have only relatively few data points, we limit the experiment to the five most frequent peptide sources. The results are shown, separated according to MHC class, in Figure~\ref{fig:multi_source_valid_base_pre}. 

\PreT{} achieves higher \AUC\, values than \BaselineT{} for all ten source/MHC combinations. We also observe that the difference is stronger for MHC I peptides than for MHC II peptides, which may be due to more room for improvement as \BaselineT{} performs worse for MHC I compared to MHC II peptides.
We also studied the performance of separate per-source models without fine-tuning. The resulting AUC values (see Supplement S6) are overall in between \BaselineT{} and \PreT{}, suggesting that the combination of both ideas, per source models and fine-tuning, are responsible for the superior performance of \PreT{}.

By generating many random permutations of the T\nobreakdash-cell response labels, we also verify that \PreT{} learns a signal that cannot be explained by random structure in the data (see Supplement S7).

\begin{figure}[ht]
  \centering
  \includegraphics[width=1.\textwidth]{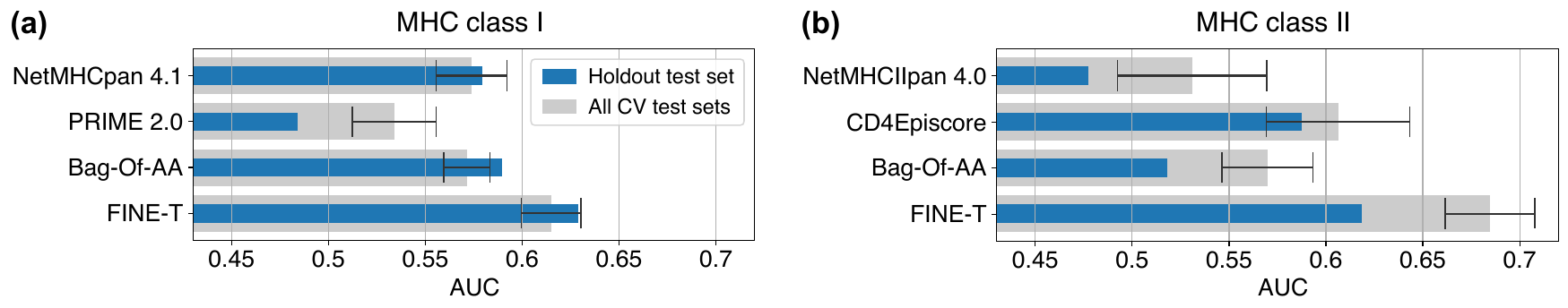}
  \caption{Test performance on human peptides presented on (a) MHC class I and (b) MHC class II. For both MHC classes, the first two rows correspond to existing models from the literature, which are trained on other data sets. Bag-of-AA baseline and \PreT{} are trained on the same data. Blue bars correspond to the evaluation on the previously completely unused test set. Grey bars show the mean AUC from the final nested cross validation.
  }
  \label{fig:human_test}
\end{figure}
\subsection{Comparison to other tools for human peptides}
T\nobreakdash-cell response predictions for peptides from human proteins are of special interest for the development of personalized cancer vaccines since neo-antigens also originate from human proteins \citep{lang2022identification}.
Thus, we focus our analysis on human peptides and compare the performance of \PreT{}, which was the best model in the validation study, to existing pre-trained models from the literature. 

NetMHCpan~4.1 and NetMHCIIpan~4.0 \citep{netmhc} have been trained to predict the presentation of peptides on MHC molecules. Despite not being trained on T\nobreakdash-cell response data, we include these models since they are commonly used in the context of personalized cancer vaccines.
PRIME~2.0~\citep{prime2_0} has been trained on T\nobreakdash-cell response data with MHC~I neo-antigens, whereas  CD4Episcore~\citep{HLA_CD4_Immunogenicity} has been trained on MHC~II T\nobreakdash-cell response data from several peptide sources.
Since the training data for all pre-trained models deviates from our training data, we also include the Bag-Of-AA baseline, trained on the same data set as \PreT{}, in the comparison. 
In contrast to our previous studies, we now train a model on the union of the first four data subsets, and evaluate on the previously unused data subset $\mathcal{D}_5$. In addition, we also carry out a full nested cross validation.
The results are shown in Figure~\ref{fig:human_test}.

For MHC~I peptides, \PreT{} achieves with a value of 0.615 the highest \AUC{} among all models. PRIME~2.0 yields the lowest performance, which is likely due to the differences in the training data.
For MHC~II peptides, \PreT{} also performs better than existing approaches, and the \AUC{} is, with a value 0.68 even slightly better than in the validation study. 
While \cite{paul2015development} have observed that combining MHC binding predictions from several alleles has a predictive value for immunogenicity, on the data set we use this applies only to MHC I (NetMHCpan 4.1) but not to MHC II (NetMHCIIpan 4.0).
The large performance difference between the Bag-Of-AA model and \PreT{} suggests that the possibility to use positional information of amino acids in the peptide sequence is indeed beneficial, and more so for MHC class II compared to MHC class I.

\section{Discussion}
In this work, we studied the effect of various transfer learning techniques on the problem of T-cell response prediction. We have observed that predicting T\nobreakdash-cell responses to peptides is a challenging problem not only due to the limited sample size, but also since the data features a multi-domain structure, which entails, as illustrated in this article, the danger of shortcut learning.

To ensure that performance estimates are not inflated by the use of shortcuts, we have defined a domain-aware evaluation scheme. We have then demonstrated that shortcut learning can be effectively reduced by applying adversarial domain adaptation \hl[in an approach dubbed \AdvT{}. This shortcut reduction did, however, not directly improve predictive performance, which is likely due to large domain differences and negative transfer between domains.

To enable positive while preventing negative transfer, we have pre-trained a transformer model on all data domains and then created fine-tuned models for individual domains, resulting in a model dubbed \PreT{}. This approach improved predictive performance considerably and performs slightly better than several existing methods, although the overall performance of all models still leaves room for improvement.

We thus conclude that the overall effect on transfer learning as a broad strategy on T-cell response prediction is mixed, with \PreT{} showing benefits, whereas \AdvT{} being not helpful. Beyond raw predictive performance, there a few other points to consider, though.

Since modeling the binding of individual T\nobreakdash-cell receptors (TCRs) and peptide:MHC complexes is difficult and does not directly capture the T\nobreakdash-cell response \citep{hudson2023can, yin2022benchmarking, fischer2020predicting}, many prediction models, including ours, assume general patterns in peptide sequences that make peptide recognition more likely in many individuals \citep{wells2020key, tcr_landscape}. 
This assumption is based on the public TCR repertoire concept of antigen-recognizing TCRs being likely shared among individuals \citep{Chen2017}.

Once more data of paired TCR and peptide-MHC complexes become available and protein structure predictions are further improved, the public TCR repertoire assumption might be relaxed. Modeling the interaction between individual TCRs and peptide-MHC complexes has the potential to improve T\nobreakdash-cell response prediction in principle \citep{nielsen2024lessons}.

In a broader context, the present study demonstrates the importance of knowing possibly hidden underlying structures in the training data for two reasons. 
First, the risk of shortcut learning is well known in domains like image recognition or natural language processing \citep{geirhos2020shortcut}. However, in biological sequence data, invalid features might be harder to discover since they can be encoded in more complex patterns spanning only specific positions in sequences, as it is for anchor positions in peptides (see Figure \ref{fig:data_domain_structure}b).
Second, knowing about the heterogeneity of a data set and viewing it as a multi-domain structure can motivate the use of transfer learning techniques. However, as we have observed with \AdvT{} and as it is also reported for other tasks such as classification of images, texts, or molecules \citep{sagawa2022extending}, transfer learning methods like adversarial domain adaptation can sometimes decrease the classification performance. Similarly, caution has to be applied in the data selection process since increasing a data set at the cost of more heterogeneity can lead to instances of negative transfer \citep{lee2023robust}.

In a similar direction it is noteworthy that to this date, all approaches for T\nobreakdash-cell response prediction use their own specific data selection criteria, differing, for example, in the selected assays, measured cytokines, or the considered peptide sources \citep{muller2023machine,prime2_0,HLA_CD4_Immunogenicity}.
We think that a benchmark data set that is well characterized in terms of its underlying structure, together with an evaluation scheme that is invariant to possible instances of shortcut learning, would be beneficial to facilitate objective development and a fair comparison of T\nobreakdash-cell response prediction methods.

\backmatter

\bmhead{Supplementary information}

There is one file with supplementary material attached to the submission.

%
%


\section*{Declarations}

\begin{itemize}
\item Funding\\
This work used the computational infrastructure of the 'Training Center Machine Learning (TCML), T\"{u}bingen' grant number 01IS17054.
\item Conflict of interest\\
No competing interest is declared.
\item Ethics approval and consent to participate\\
Not applicable
\item Consent for publication\\
Not applicable
\item Data availability\\
Data is available under \href{https://github.com/JosuaStadelmaier/T-cell-response-prediction}{github.com/JosuaStadelmaier/T-cell-response-prediction}.
\item Materials availability\\
Not applicable
\item Code availability\\
Source code and trained models are available under \href{https://github.com/JosuaStadelmaier/T-cell-response-prediction}{github.com/JosuaStadelmaier/T-cell-response-prediction}.
\item Author contribution\\
J.S., B.M., and R.E conceived the experiment(s), J.S. implemented the software, J.S. conducted the experiment(s), J.S., B.M., and R.E. analysed the results.  J.S., B.M. and R.E. wrote and reviewed the manuscript.

\end{itemize}



\bibliography{document.bib}

\end{document}


\maketitle

\section{Joint distribution of MHC alleles and peptide sources}
The following heatmap shows the most frequent peptide sources and a representative subset of MHC alleles. As mentioned in Section 2.2, the patterns in the heatmap reveal that the joint distribution of MHC alleles and peptide sources is not uniform.

\begin{center}
\includegraphics[width=0.9\textwidth]{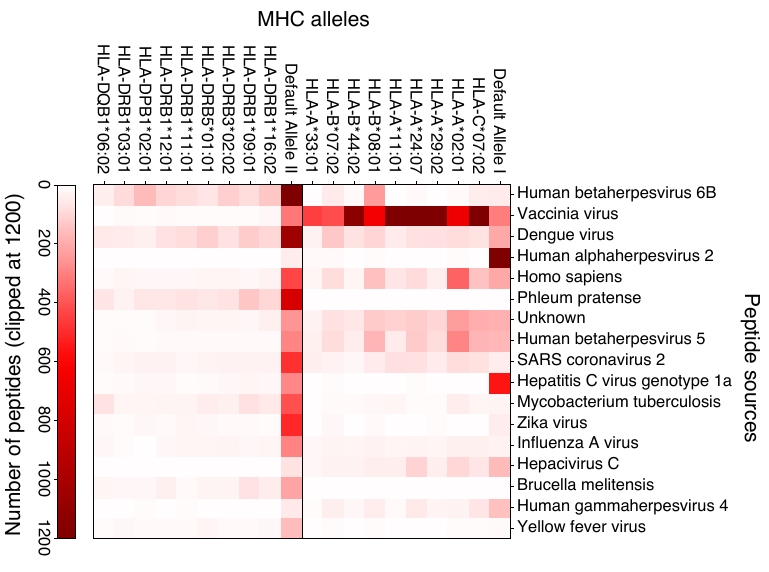}
\end{center}

\clearpage

\section{Domain-aware evaluation}
We ensure that our evaluation is not influenced by the use of MHC allele-based or peptide source-based shortcuts by evaluating our models on subsets of the data containing only one MHC allele and one peptide source.
Shortcuts in the form of predicting the majority class of a peptide source or an MHC allele do not provide predictive power within such subsets.
The performance on these subsets thus represents what a model has learned beyond the shortcuts.

When summarizing the per-subset results to an overall score, we ensure that every peptide has the same influence on the overall score, even if it occurs in several subsets as a consequence of being presented on several MHC alleles.

To describe our domain-aware evaluation setup in more detail, we introduce the following formalization: Let $(G_k)_{k=1,\dots,n} \, (n \leq \#\text{alleles} \cdot \#\text{sources})$ be the list of peptide groups.
Each peptide group $G_k$ corresponds to an allele-source combination. A group $G_k$ contains indexes of peptides from only one source and one allele.
For the i-th peptide in our data set, we denote the T-cell response label as $y_i$, the prediction as $\hat{y}_i$ and corresponding lists for a peptide group $G$ as $\bm{y}_G = (y_i)_{i \in G}$ and $\bm{\hat{y}}_G = (\hat{y}_i)_{i \in G}$.

To achieve that every peptide has the same influence on the overall AUC score, despite being present in varying numbers of groups, we define peptide weights $w_i$ to be the inverse of the number of groups the peptide is part of: $w_i = \frac{1}{|\{G_k \mid i \in G_k \}|}$. We define the total weight of peptides associated with a group as $w_G = \sum_{i \in G} w_i$. Analogously to the labels and predictions, we denote the list of peptide weights for a group $G$ as $\bm{w}_G = (w_i)_{i \in G}$.

The AUC score that is adjusted for both source and allele-based shortcuts can now be defined as follows:
\begin{equation}
    \text{AUC} = \sum_{k=1}^n \frac{w_{G_k}}{\sum_{k=1}^n w_{G_k}} \cdot \texttt{roc\_auc\_score}(\bm{y}_{G_k}, \bm{\hat{y}}_{G_k}, \text{sample\_weight}= \bm{w}_{G_k})
\end{equation}

We use the \texttt{roc\_auc\_score} function from Scikit-learn \citep{scikit-learn}. Weighting the per-group AUC scores with the normalized per-group weights accounts for both varying group sizes and for peptides being part of varying numbers of groups.

To obtain the "allele adjusted" results from Figure 3, we define the peptide groups based on MHC alleles only, s.t. each group contains peptides from only one allele. Then we apply the same weighting scheme as above. This setting only adjusts for shortcuts based on MHC alleles. Analogously, we can also obtain the "source adjusted" results from Figure 3.

\clearpage

\section{Implementation and hyperparameter selection}
For implementing the experiments, we use PyTorch and its built-in implementation of the transformer encoder.
We perform gradient-based optimization using the stochastic optimizer Adam \citep{adam}.
Besides regularization with dropout, we apply early stopping based on the validation performance with a maximum of 250 epochs.

To reduce the uncertainty of performance estimates resulting from the random initialization of network weights, sampling of mini-batches, and dropout, we repeat every experiment with different random seeds and then average the performance estimates.

We use a grid search combined with cross-validation to select suitable values for hyperparameters.
Since exhaustively testing every hyperparameter combination of the transformer models in each experiment is computationally expensive, we perform an initial exploration of hyperparameters on all peptide sources and MHC alleles combined and then use smaller, more informed hyperparameter ranges for the individual experiments (see Table \ref{tab:transformer_hyperparams}). Unless otherwise stated, we set the batch size to 500. The hyperparameter ranges for the Bag-Of-AA model are shown in Table \ref{tab:bag_of_aa_hyperparams}.

The validation performance results from Section 3.1 to Section 3.3 are obtained by running the cross-validation on the first four data partitions $\mathcal{D}_1,\dots,\mathcal{D}_{4}$.
The test results with standard error estimates from Section 3.4 are obtained by a nested cross-validation on all data partitions $\mathcal{D}_1,\dots,\mathcal{D}_{5}$.

   \begin{table}[ht]
      \caption{Hyperparameter ranges for the transformer models}
      \label{tab:transformer_hyperparams}
      \centering
         \begin{tabular}{cccccc}
           \toprule
      & Learning &  & Embedding & Attn. & Attn.\\
    Experiment & rate & Dropout & dimension $d$ & layers & heads\\
    \toprule

    Initial exploration	 & \{0.001,  0.01, 0.05\} & \{0, 0.1, 0.2, 0.3\}	& \{16, 32, 64, 96\}& \{1, 2\}   & \{4, 8, 16\}  \\
  \midrule
 
    Sec. 3.1 / Fig. 3 & 0.005 & 0.2	& 64   & 1    & 8  \\
    \multicolumn{6}{l}{ADA-T: Select $\lambda \in $ \{2, 4, \dots, 22\} to minimize $\mid$ source-adjusted AUC - AUC w/o adjustments $\mid$.}\\
  \midrule

    Sec. 3.2 / Supp. 2 & 0.01 & \{0, 0.1, 0.3\} & 32   & 1    & 8  \\
    \multicolumn{6}{l}{Batch size set to 100 due to smaller data sets. Max epochs set to 100 for efficiency.}\\
\midrule

    Sec. 3.3 / Supp. 3 & 0.01 & \{0.1, 0.2, 0.3, 0.4\} & \{32, 48\}   & 1    & 16  \\

\midrule

    Sec. 3.4 FINE-T I & 0.01 & \{0.1, 0.2\} & \{32, 48\}   & 1    & 16  \\
    Sec. 3.4 FINE-T II & 0.01 & \{0.2, 0.3\} & \{32, 48\}   & 1    & 16  \\

    \bottomrule
    \end{tabular}

    \end{table}

\begin{table}[ht]
      \caption{Hyperparameter ranges for the Bag-Of-AA baseline}
      \label{tab:bag_of_aa_hyperparams}
      \centering
         \begin{tabular}{cccccc}
           \toprule
    Experiment & Learning rate & Dropout & Hidden layer dimension\\
    \toprule

 
    Sec. 3.4 Bag-Of-AA I / II & \{0.01, 0.1, 0.2\} & \{0.1, 0.2\}	& \{64, 96, 128\} \\

    \bottomrule
    \end{tabular}

    \end{table}

\subsection{NetMHCpan 4.1 and NetMHCIIpan 4.0 predictions}
We include NetMHCpan 4.1 EL and NetMHCIIpan 4.0 EL to in our final test evaluation of T-cell response predictions for human peptides. With this, we aim to capture how much of the predictive power of the T-cell response models can be achieved with existing models of MHC presentation.

\cite{paul2015development} have observed that combining the MHC binding predictions for several alleles for a given peptide can provide immunogenicity estimates.
Similar to the approach by \cite{paul2015development},
we take the minimum of percentile rank scores over all MHC I or MHC II alleles in our data set to obtain estimates or T-cell responses.
\clearpage
\section{Shortcut learning influence on prediction scores}
Similar to Figure 3b, we show the distributions of prediction scores separated by the ground truth label for both the BASE-T model and the ADA-T model.
In contrast to Figure 3b, each peptide source is shown as a separate row of histograms.

\begin{center}
    \includegraphics[width=0.85\textwidth]{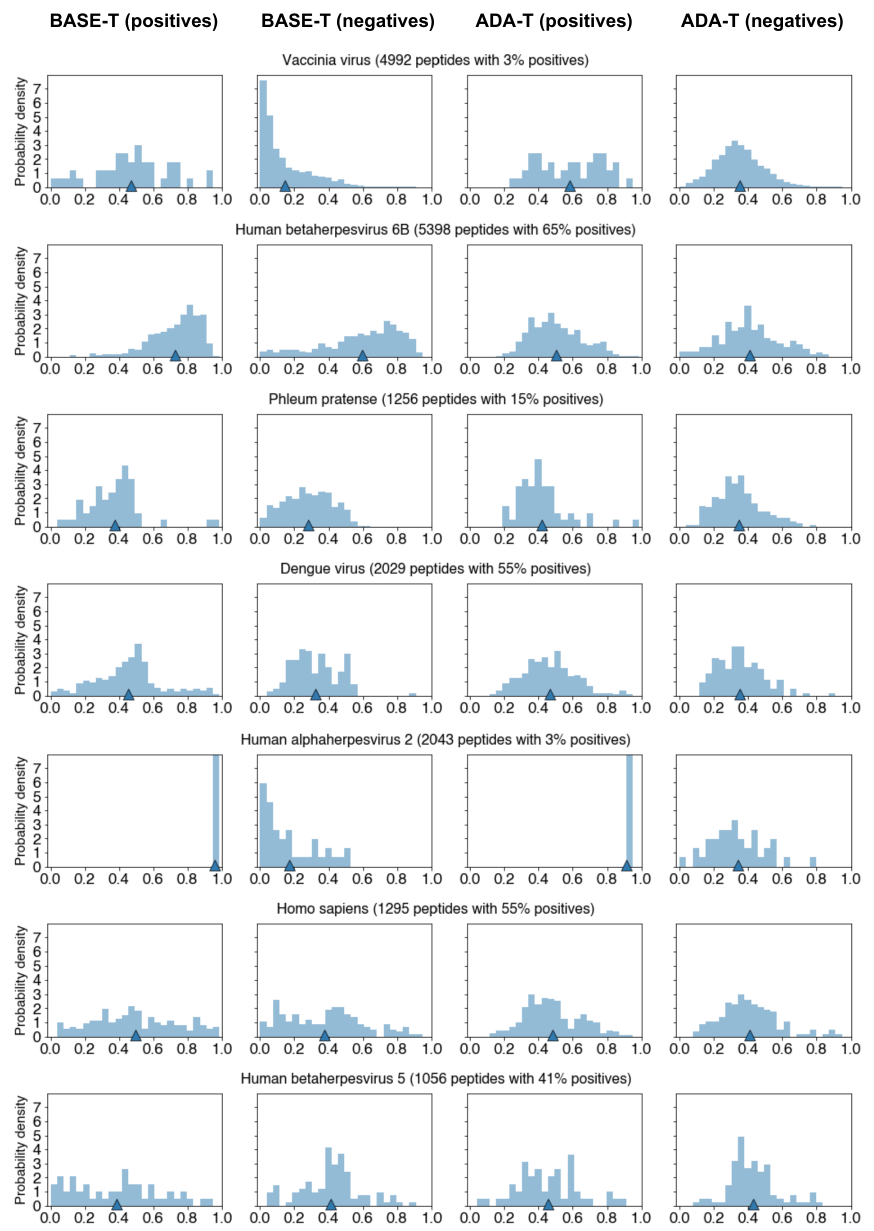}
\end{center}

\clearpage

\section{Negative transfer between domains}

As described in Section 3.2, we compare \BaselineT{}, being trained all peptide sources and MHC alleles (multi-domain setting), with a combination of \BaselineT{} models that eliminate transfer between sources and alleles respectively.

The following plot shows the validation performance of these three settings.
The observation that per-source models and to a smaller degree per-allele models lead to more accurate predictions than the multi-domain model implies that the multi-domain model suffers from negative transfer between peptide sources and to a smaller degree between MHC alleles.

\begin{center}
  \includegraphics[width=0.6\textwidth]{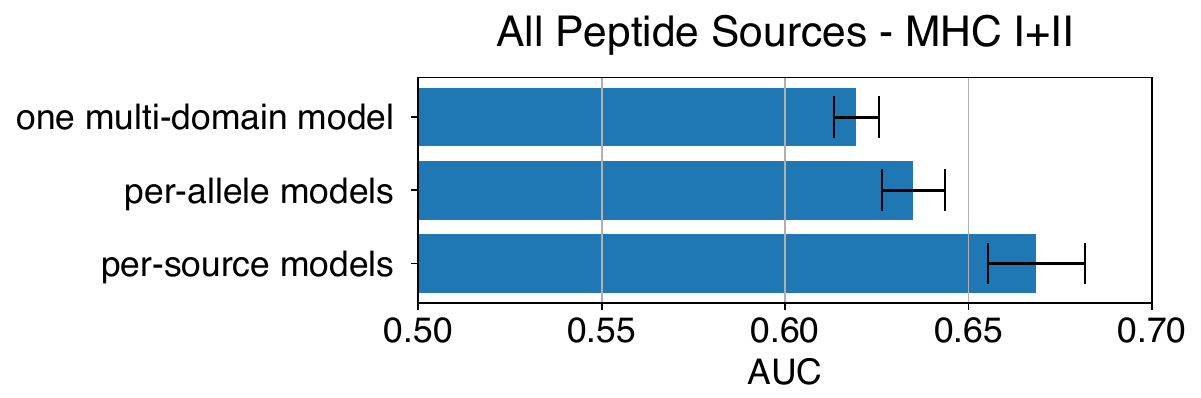}
\end{center}

\clearpage

\section{Ablation study for \PreT{}}
To investigate the contribution of fine-tuning and the benefit of mitigating negative transfer, we show the validation performance of transformer models in different transfer learning settings for several peptide sources.

Results for MHC I are shown in \captionbf{(a)} and for MHC II in \captionbf{(b)}. For each MHC class, the five most frequent peptide sources are selected. Peptide sources with only positives or only negatives in one of the data partitions are excluded.

This analysis shows that both mitigating negative transfer (multi-domain vs. per-source) and fine-tuning (\PreT{} vs. per-source) consistently improve the predictions.

\vspace{0.5cm}

\begin{center}
    \includegraphics[width=0.64\textwidth]{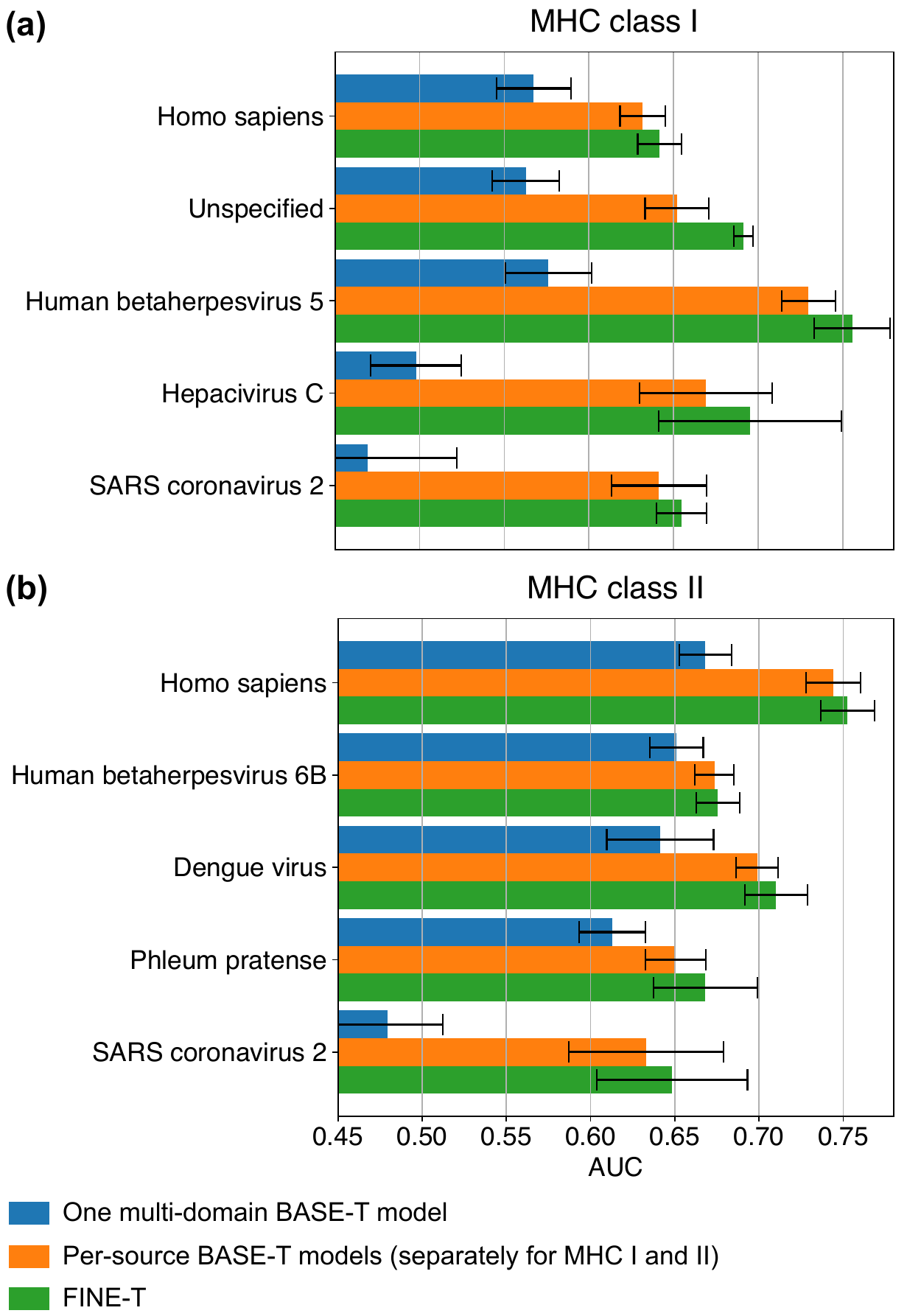}
\end{center}

\clearpage

\section{Permutation experiment with \PreT{}}

We randomly permute the labels of training and test data sets 200 times to obtain the performance distribution of the \PreT{} model when only random patterns are picked up. Results for MHC I are shown in \captionbf{(a)} and for MHC II in \captionbf{(b)} in the following figure.

We apply the permutation always within the training data set and within the test data set and apply the same training and evaluation procedure as for obtaining the test performance in Figure 5 (grey bars).

The red arrow indicates the performance of the \PreT{} model when trained and evaluated on the data sets without permutation. Consequently the red arrows show the same performance as the grey bars for \PreT{} in Figure 5.
Since the red arrows lie outside of the performance distribution of the permutation experiment, the performance of \PreT{} cannot be explained by learning random patterns.

\vspace{0.5cm}

\begin{center}
    \includegraphics[width=0.6\textwidth]{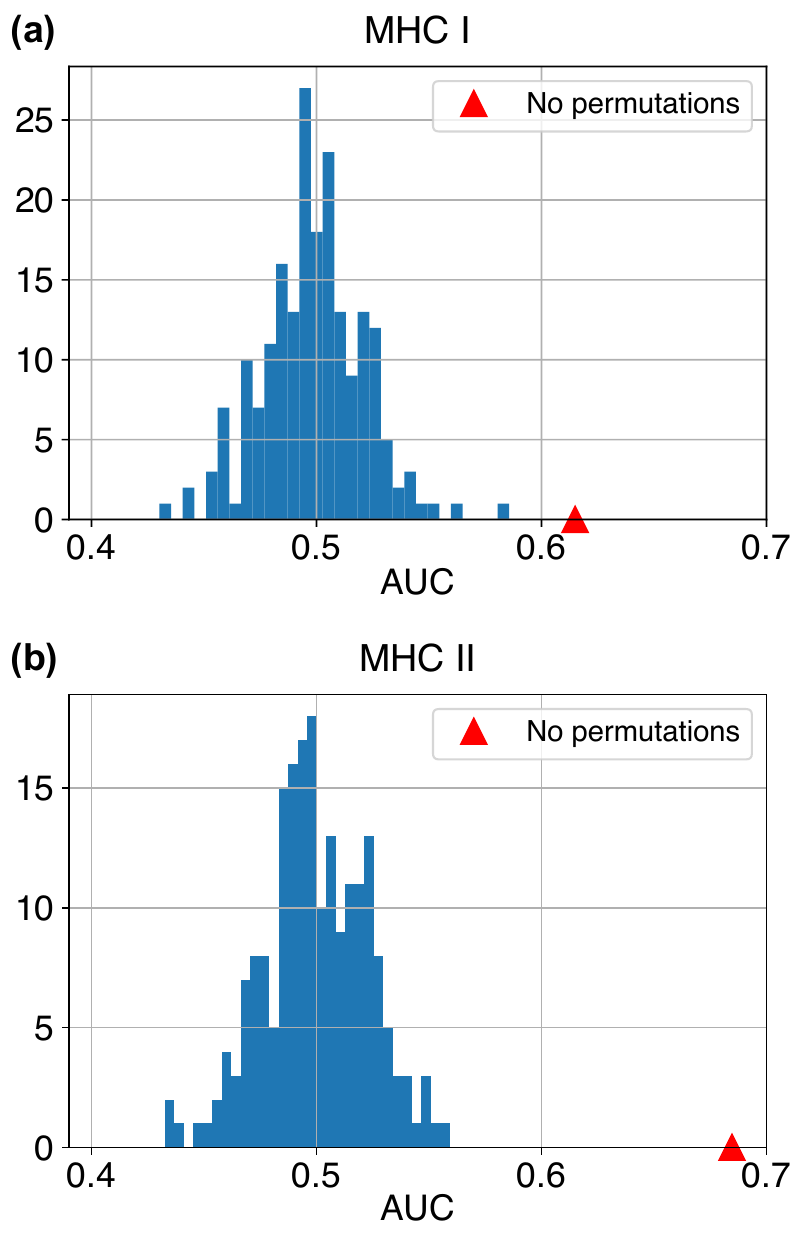}
\end{center}

\clearpage

\bibliographystyle{natbib}

\bibliography{document}